\documentclass[reprint,amsmath,amssymb,aps,superscriptaddress,prd]{revtex4-2}
\usepackage{hyperref}
\usepackage[utf8]{inputenc}
\usepackage{parskip}
\usepackage{graphicx}
\usepackage{array}
\usepackage{siunitx}
\usepackage{xcolor}

\newcommand*{\dif}{\text{d}}
\graphicspath{{./figs/}}

\begin{document}

\title{Rapid onset of the 21-cm signal suggests a preferred\\mass range for dark matter particle}

\author{Venno Vipp}
\email{venno.vipp@kbfi.ee}
\affiliation{Institute of Physics, University of Tartu, W. Ostwaldi Str 1, 50411 Tartu, Estonia}
\affiliation{NICPB, Rävala 10, 10143 Tallinn, Estonia}
\author{Andi Hektor}
\email{andi.hektor@cern.ch}
\affiliation{NICPB, Rävala 10, 10143 Tallinn, Estonia}
\author{Gert Hütsi}
\email{gert.hutsi@to.ee}
\affiliation{NICPB, Rävala 10, 10143 Tallinn, Estonia}

\date{\today}

\begin{abstract}
We are approaching a new era to probe the 21-cm neutral hydrogen signal from the period of cosmic dawn. This signal offers a unique window to the virgin Universe, \emph{e.g.}, to study dark matter models with different small-scale behaviours. The EDGES collaboration has recently published the first results of the global 21-cm spectrum. We demonstrate that such a signal can be used to set, unlike most observations concerning dark matter, both lower \emph{and} upper limits for the mass of dark matter particles. We study the 21-cm signal resulting from a simple warm dark matter model with a sharp-$k$ window function calibrated for high redshifts. We tie the PopIII star formation to Lyman-alpha and radio background production. Using MCMC to sample the parameter space we find that to match the EDGES signal, a warm dark matter particle must have a mass of $7.3^{+1.6}_{-3.3}$ keV at 68\% confidence interval. This translates to $2.2^{+1.4}_{-1.7} \times 10^{-20}$ eV for fuzzy dark matter and $63^{+19}_{-35}$ keV for Dodelson-Widrow sterile neutrinos. Cold dark matter is unable to reproduce the signal due to its slow structure growth.
\end{abstract}

\maketitle


\section{Introduction}


The current concordance cosmology, $\Lambda$CDM~\cite{ostriker_cosmic_1995,bahcall_cosmic_1999}, has been around for more than two decades. The model has been amazingly robust, being able to accommodate most of the new observational data~\cite{riess_observational_1998,perlmutter_measurements_1999,de_bernardis_flat_2000,hanany_maxima-1_2000,spergel_first_2003, tegmark_cosmological_2004,cole_2df_2005,aghanim_planck_2020}. Despite this enormous success some cracks have started to emerge, the most notable being the Hubble tension~\cite{riess_24_2016}. In addition, there have been several widely acknowledged problems related to structure formation at smaller scales: namely, the core-cusp~\cite{moore_evidence_1994}, the missing satellites~\cite{klypin_where_1999,moore_dark_1999} and too big to fail~\cite{boylan-kolchin_too_2011} problems, to mention a few most often discussed. 

To address these problems with $\Lambda$CDM several alternatives which have suppressed power at small scales, have been proposed, \emph{e.g.}, warm dark matter (WDM)~\cite{colin_substructure_2000,bode_halo_2001, viel_warm_2013}, fuzzy dark matter (FDM)~\cite{hu_cold_2000, marsh_axion_2016, hui_ultralight_2017} or self-interacting dark matter~\cite{spergel_observational_2000, vogelsberger_subhaloes_2012, rocha_cosmological_2013}.

The concordance $\Lambda$CDM model is based on high-redshift cosmological probes (Big Bang nucleosyntesis, cosmic microwave background spectral distortions and anisotropies) and low-redshift measurables (large-scale structure, type Ia supernovae). Recently a new important probe at intermediate redshifts has emerged: the 21-cm absorption feature from the cosmic dawn ($z\sim 20$) as seen by the EDGES experiment~\cite{bowman_absorption_2018}. If true, this has enormous consequences; it allows us for the first time to probe the onset of structure formation, and helps us test models with small-scale behaviour different from the $\Lambda$CDM model.

Many authors have attempted to explain the EDGES signal using various mechanisms~\cite{feng_enhanced_2018,ewall-wice_modeling_2018,fraser_edges_2018,lopez-honorez_dark_2019,chatterjee_ruling_2019,natwariya_edges_2020}, the most popular among them being those which lower the gas temperature via baryon-dark matter interactions, \emph{e.g.}, millicharged dark matter models~\cite{barkana_possible_2018, liu_too_2018, munoz_insights_2018}. However, most of the works have used the spectral location and/or amplitude of the EDGES absorption feature in their modeling, rather than the rate of its onset. Namely, the EDGES measurement suggests a very sudden decrease of spin temperature, which implies a rapid growth in Lyman-alpha radiation (Ly$\alpha$) production. In a standard $\Lambda$CDM scenario, this is very difficult to achieve as structure formation proceeds too mildly to result in the rapidly-changing EDGES spectrum. Kaurov \emph{et al} studied the implications of the rapid onset of the signal in \cite{kaurov_implication_2018}. They estimated that such a rapid signal would require star formation to occur mostly in the exponential tail of the halo mass function (HMF) and accordingly adopted a model where the minimum halo mass for star formation was $ \sim 10^{9.5} \, M_\odot$. We find that the star formation history of such a small-scale suppressed model is qualitatively identical to warm dark matter models.

It is important to stress that the rapid change in antenna brightness temperature ($T_{21}$) as observed by EDGES carries great significance: only the dramatic sharpness of the signal allows it to be successfully extracted from the (presumed) smooth astrophysical and atmospheric foreground contributions. Thus, the rapid onset of this signal, within the redshift range $\sim 18<z<23$, is a particular focus of this study. While many studies of non-cold dark matter can give lower limits for the particle mass (thus upper limits for the amount of suppression at small scales) we show that the EDGES signal, more specifically its onset, provides both \emph{upper} and \emph{lower} bounds. 

In this paper we study small-scale suppressed models within the context of the EDGES results. We apply Markov Chain Monte Carlo (MCMC) with the emcee package~\cite{foreman-mackey_emcee_2013} to model WDM and to find parameter combinations capable of fitting the rapid onset of the EDGES signal. Though we run the calculation only on WDM the results can be mapped (equations \eqref{eq:wdm_to_fdm} and  \eqref{eq:wdm_to_nu}) to other dark matter (DM) models with suppressed small scale power, namely, FDM and sterile neutrinos. We include an extra radio background and the Ly$\alpha$ coupling of spin temperature $T_S$ to the kinetic temperature of gas $T_K$. We focus on the onset of the signal and not the overall shape or central position, and therefore X-ray, Ly$\alpha$, and Compton heating do not have enough time to proceed. So heating plays a minor role in our modelling process. For completeness, and to illustrate the resulting full signal shape, we include a simple X-ray heating model. We take all the baseline cosmological parameters as presented by the Planck collaboration \cite{aghanim_planck_2020}.

Our results show that to match the EDGES signal, the mass of WDM particle must be in the range of $7.3^{+1.6}_{-3.3}$~keV (68\% CL). For fuzzy DM this translates to $2.2^{+1.4}_{-1.7} \times 10^{-20}$~eV and for Dodelson-Widrow sterile neutrinos to $63^{+19}_{-35}$~keV. Surprisingly, CDM cannot reproduce the signal due to its too slow structure growth, which cannot produce the rapidly increasing onset of the EDGES spectrum.

The paper is organized as follows. In section \ref{sec:method} we outline the calculation of all the relevant parameters: $T_{21}$ and Ly$\alpha$ intensity from EDGES along with their errors (in~\ref{ssec:lya_edges}), star formation rates and Ly$\alpha$ intensity from the WDM model (in~\ref{ssec:sfrd}). We explain the adding of extra radio background and X-ray heating (in~\ref{ssec:radio_xray}). In section \ref{sec:results} we present our results and in section \ref{sec:conclusions} draw our conclusions.


\section{Method}\label{sec:method}


In this section we present our model for calculating matter power spectra, halo mass functions, star formation rates and radiation intensities from the EDGES signal and from the WDM model.


\subsection{Ly\texorpdfstring{$\boldsymbol{\alpha}$}{a} radiation according to EDGES} \label{ssec:lya_edges}


The EDGES collaboration measured the brightness temperature of the 21-cm global signal. This is determined by the ratio of the temperature of background radiation $T_R$ and the spin temperature $T_S$ \cite{madau_21-cm_1997} 
\begin{gather} \label{eq:antenna_temp}
    T_{21} \simeq 26.8 \, \text{mK} \dfrac{\Omega_b h}{0.0327} \sqrt{ \dfrac{0.307}{\Omega_m} \dfrac{1+z}{10}} \left( 1-\dfrac{T_R}{T_S} \right)
\end{gather}
where $\Omega_m$ and $\Omega_b$ are the density parameters for matter and baryonic matter respectively and $h$ is the Hubble constant in units of 100 km s$^{-1}$ Mpc$^{-1}$. In the absence of other sources, the background temperature is just the temperature of the CMB at that redshift $T_R(z) = 2.725(1+z)$ K~\cite{fixsen_temperature_2009}. $T_S$ is determined by the relative abundances of HI atoms in their ground and excited states of the 21-cm spin-flip transition. It can also be calculated as a weighed average \cite{field_excitation_1958, barkana_rise_2016}
\begin{equation}\label{eq:ts}
    T_S^{-1} = \dfrac{x_\alpha T_K^{-1} + T_R^{-1}}{x_\alpha + 1},  \quad  x_\alpha = S_\alpha \dfrac{J_\alpha}{J_\alpha^c}, \\
\end{equation}
where $T_K$ is the kinetic temperature of gas, $J_\alpha$ is the Ly$\alpha$ background intensity, \mbox{$J_\alpha^c = \SI{9e-23} (1+z) $} $\mathrm{\, ergs \, cm^{-2} \, s^{-1} \,Hz^{-1} \,sr^{-1}}$~\cite{ciardi_probing_2003} the critical Ly$\alpha$ intensity and the correction factor $S_\alpha$ is \cite{barkana_rise_2016}
\begin{gather} 
S_\alpha = \exp\left[ {-\SI{0.0128}{} \left(\dfrac{\tau_{GP}}{T_K^2}\right)^{1/3}} \right], \label{def_S_alpha}\\
\tau_{GP} = \SI{6.6e5}{} \left( \dfrac{\Omega_bh}{\SI{0.0327}{}} \right) \sqrt{\left( \dfrac{\Omega_m}{\SI{0.307}{}} \right)^{-1} \left( \dfrac{1+z}{10}\right)}.\label{def_tau_GP}
\end{gather}
Combining the above equations one can express $J_\alpha$ as
\begin{equation}\label{eq:jaedges}
    J_\alpha = \dfrac{J_\alpha^c}{S_\alpha} \dfrac{T_R^{-1} -T_S^{-1}}{T_S^{-1} - T_K^{-1}},
\end{equation}
in which $T_S$ can be calculated using equation \eqref{eq:antenna_temp}.

To calculate $J_\alpha$ accordingly one must make assumptions about $T_K$ and $T_R$. Since the spin temperature is a weighted average of $T_K$ and $T_R$, its value must always be between the two. But, in the standard cosmological case with no extra radiation background or cooling of gas, calculating the spin temperature from equation \eqref{eq:antenna_temp} from the EDGES signal leads to it dropping below both $T_K$ and $T_R$ at $z\sim 20$, simultaneously pushing $J_\alpha$ to infinity at the point of equality (see figure \ref{fig:sfrdcomp}). The cause of this is the high amplitude of the signal. This has led many to look at models that either reduce the temperature of gas (e.g., see~\cite{barkana_possible_2018,liu_too_2018,munoz_insights_2018,natwariya_edges_2020}) or add extra radio background (\emph{e.g.}, see~\cite{feng_enhanced_2018,ewall-wice_modeling_2018,fraser_edges_2018,chatterjee_ruling_2019}) and others to disregard the amplitude and focus only on the timing of the signal~\cite{schauer_constraining_2019, boyarsky_21-cm_2019, leo_constraining_2020}.

We find a good approximation of the $J_\alpha$ implied by the onset of EDGES to be a powerlaw
\begin{equation} \label{eq:ja_powerlaw}
    J_\alpha(z) = J_{\alpha, 0} \left( \dfrac{1+z}{1+z_0} \right)^{n(z)},
\end{equation}
where $z_0=24$ and power index $n(z) \simeq -1.67 z - 10.45$. This is in the case of no extra radio background and unchanged kinetic temperature history. It is important to note that for $J_\alpha$ parametrized in this way $n(z)$ does not depend significantly on the amplitude of the $T_{21}$ signal, only $J_{\alpha, 0}$ does (see figure \ref{fig:sfrdcomp}). So while some values of the EDGES signal amplitude give unphysical results, because they imply $T_S<T_K<T_R$, the values of spectral index $n(z)$ are similar for all amplitude values. This means that the fast growth of Ly$\alpha$ radiation in the cosmic environment is not the result of an unusually large signal amplitude, but rather the overall shape of the signal.

We also find the error bounds for EDGES related values using the confidence intervals (CIs) given by the EDGES collaboration. The 99\% CI provided by \cite{bowman_absorption_2018} are asymmetric and the corresponding parameter distributions are not known. To approximate errors we take the four parameters $A, \nu_0, w$ and $\tau$ to be uncorrelated and find the best probability distributions for them by fitting the means and 99\% CIs. Finally, we generate a large amount of new signal parameters from these distributions and use them to find the 1$\sigma$ and 2$\sigma$ error bounds for $T_{21}$ and the corresponding Ly$\alpha$ intensity in the case of EDGES results. Figures \ref{fig:sfrdcomp} and \ref{fig:t21comp} show the calculated bounds.


\subsection{Star formation and Ly\texorpdfstring{$\boldsymbol{\alpha}$}{a} intensity in DM models}\label{ssec:sfrd}


We calculate the matter power spectrum of CDM using the fitting formulae from from \cite{eisenstein_baryonic_1998}. Beyond CDM one can simply modify the CDM power spectrum with the corresponding transfer function,
\begin{equation}
    P_{\mathrm{DM}}(k,z,m) = P_{\mathrm{CDM}}(k,z) \, T_{\mathrm{DM}}^2(k,m).
\end{equation}
The transfer function for WDM can be approximated as~\cite{bode_halo_2001},
\begin{equation}
    T_{\mathrm{WDM}}(k,m) \simeq [1+ (\alpha k)^{2\nu} ]^{-5/\nu}, 
\end{equation}
where $\nu = 1.12$ \cite{viel_constraining_2005} and 
\begin{equation}
    \alpha \simeq 0.049 \left( \dfrac{m_{x}}{1\mathrm{ keV}} \right)^{-1.11} \left( \dfrac{\Omega_{x}}{0.25} \right)^{0.11} \left( \dfrac{h}{0.7} \right)^{1.22} h^{-1} \mathrm{ Mpc}.
\end{equation}
Here $m_x$ is the mass of the WDM particle and $\Omega_x$ the density parameter for WDM. 

\begin{figure}
    \includegraphics[width=0.48\textwidth]{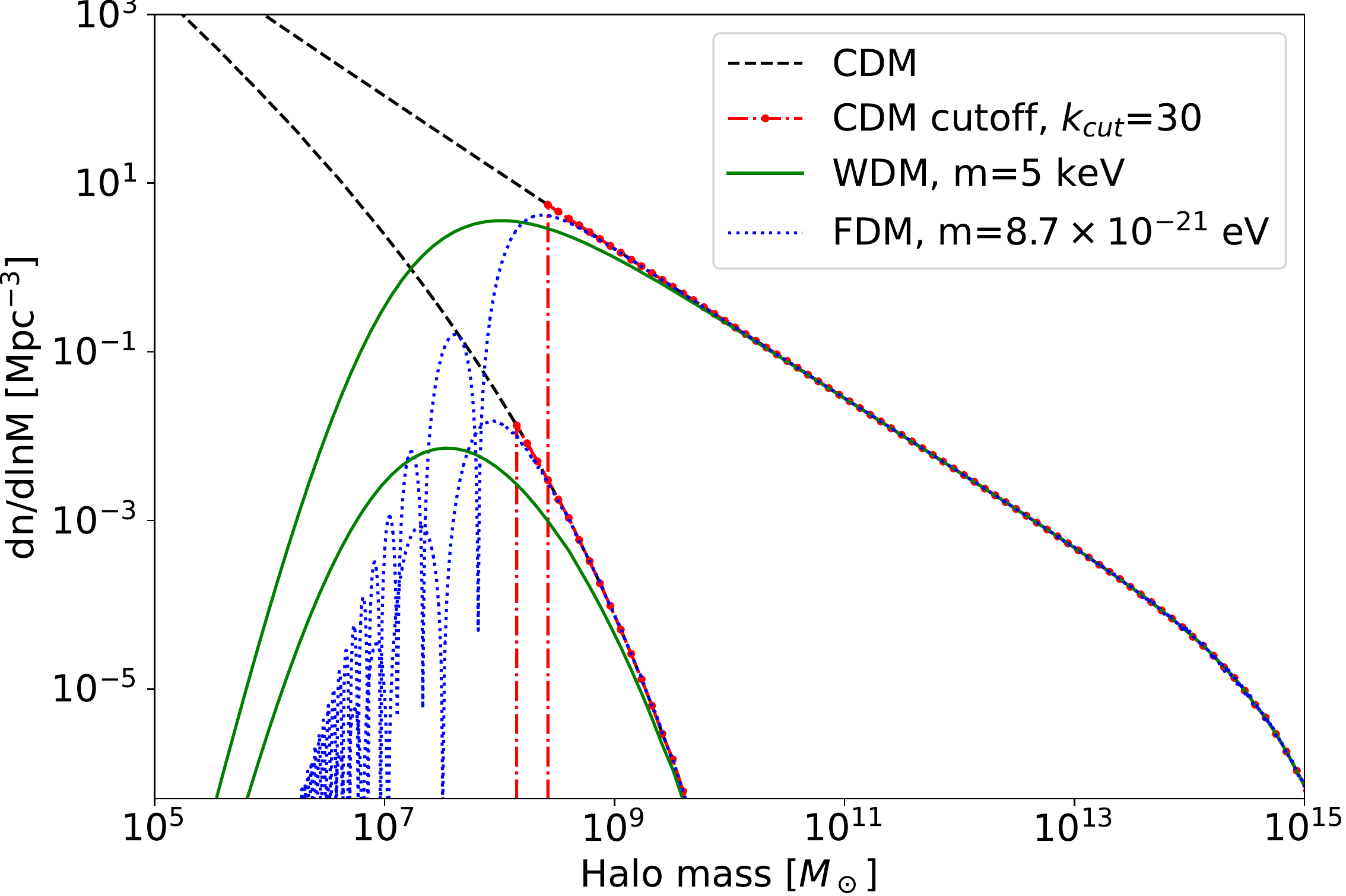}
    \caption{Comparison of HMFs for CDM (black dashed), WDM with 5 keV mass (green solid), FDM with $8.7 \times 10^{-21}$ eV mass (blue dotted), and CDM with a cutoff in its power spectrum at $k_{\mathrm{cut}}=30 \, h/\mathrm{Mpc}$ (red dot-dashed). The lines are calculated at redshifts 0 (upper curves) and 20 (lower curves). Here we use the Schneider formalism~\cite{schneider_halo_2013} with a sharp-$k$ window function for models other than CDM. The masses and cutoff scale are chosen to result in approximately equal Ly$\alpha$ output.}
    \label{fig:hmfcomp}
\end{figure}

There are formulas which roughly map the masses of particles in different DM models directly onto each other. For example the mapping between Dodelson-Widrow sterile neutrinos~\cite{dodelson_sterile_1994} and WDM is~\cite{colombi_large_1996}
\begin{equation} \label{eq:wdm_to_nu}
    m_{\nu} \simeq 4.43 \mathrm{keV} \left( \dfrac{m_{x}}{1 \mathrm{keV}} \right)^{4/3} \left( \dfrac{0.25 \cdot 0.7^2}{\omega_x} \right)^{1/3}, 
\end{equation}
where $\omega_x = \Omega_x h^2$ is present day energy density of DM. The corresponding mapping between FDM and WDM is~\cite{marsh_axion_2016}
\begin{equation} \label{eq:wdm_to_fdm}
    m_{\mathrm{FDM}} \simeq 1.55 \left( \dfrac{ m_{x} }{\mathrm{keV}} \right)^{2.5}  \cdot 10^{-22} \mathrm{eV}.
\end{equation}
We use this equivalency between models to run calculations only for WDM and translate the resulting mass bounds into their equivalent FDM and sterile neutrino masses. 
In addition, since these non-CDM models are all modifications of CDM with some manner of cutoff in their power spectra, we also consider a CDM model with a simple sharp cutoff in its power spectrum above some wavenumber $k_{\mathrm{cut}}$ (see figure \ref{fig:hmfcomp}), phenomenologically similar to the model studied in~\cite{kaurov_implication_2018}. This model would be as sharp as possible in its power spectrum and HMF and one might accordingly expect it to have a similarly sharp star formation history.

To find the HMFs of our models at very high redshifts we use the model detailed in \cite{schneider_halo_2013} and \cite{schneider_structure_2015}. The model is calibrated with high resolution numerical simulations and uses a sharp-$k$ window function to accurately reproduce WDM HMFs even at higher redshifts where the extended Press-Schechter with a tophat filter is less accurate. The model was also used in \cite{schneider_constraining_2018} to put bounds on different DM models using the timing of the EDGES signal. In figure \ref{fig:hmfcomp} we show HMFs for CDM, WDM, FDM and the model of CDM with a cutoff. Here the relevant parameters are chosen so as to result in approximately equal Ly$\alpha$ output. The sterile neutrino model with appropriately adjusted particle mass has a HMF very similar to WDM~\cite{colombi_large_1996, bode_halo_2001}.

We follow the general Ly$\alpha$ radiation model outlined in \cite{barkana_detecting_2005}. The angle averaged spectral intensity generated by sources at some redshift $z$ is \cite{barkana_detecting_2005}
\begin{gather}
J_\alpha = \dfrac{(1+z)^2 }{4\pi} \sum_{n=2}^{n_{\mathrm{max}}} f_{\mathrm{rec}}(n) \int_{z}^{z_{\mathrm{max}}} \dfrac{c \, \epsilon(\nu^\prime, z^\prime)}{H(z^\prime)} \dif z^\prime
\end{gather} 
where $c$ is the speed of light, $\epsilon(\nu^\prime)$ is the emissivity of stars, and $H(z)$ the Hubble function. This takes into account photons from higher Ly$n$ lines, with emitted frequency $\nu^\prime = \nu (1+z^\prime)(1+z)$, cascading to Ly$\alpha$, with probabilities $f_{\mathrm{rec}}(n)$ (for which values are tabulated in \cite{pritchard_descending_2006}) showing the fraction of photons recycled. In order to cascade in this way these photons must originate from within a redshift of
\begin{equation}
    z_{\mathrm{max}} = \dfrac{1- (n+1)^{-2}}{1-n^{-2}} (1+z) -1 
\end{equation}
and $n_{\mathrm{max}}$ is taken to be $\sim 23$. $\epsilon(\nu, z)$ depends on the star formation rate density (SFRD) $\dot{\rho}_\star(z)$ of a given model and the emissivity model $\epsilon_b(\nu)$ for those stars:
\begin{gather}
\epsilon(\nu, z) = \epsilon_b(\nu) \dot{\rho}_\star(z) = \epsilon_b(\nu) f_\star \rho_b^0 \dfrac{\mathrm{d}f_{\mathrm{coll}}(z)}{\mathrm{d}t}   ,
\end{gather}
where $f_\star$ is star formation efficiency (SFE) and $\rho^0_b$ is the mean baryon density at $z=0$. In truth $f_\star$ can be considered to contain factors besides SFE, such as Ly$\alpha$ escape fraction, and is likely time-dependent. Because there is uncertainty in all of the values of these factors at such high redshifts we simply lump them together into one variable and prefer lower values in our analysis. In MCMC calculations we set $f_\star$ an upper bound of 0.15.

\begin{figure}
    \centering
    \includegraphics[width=0.48\textwidth]{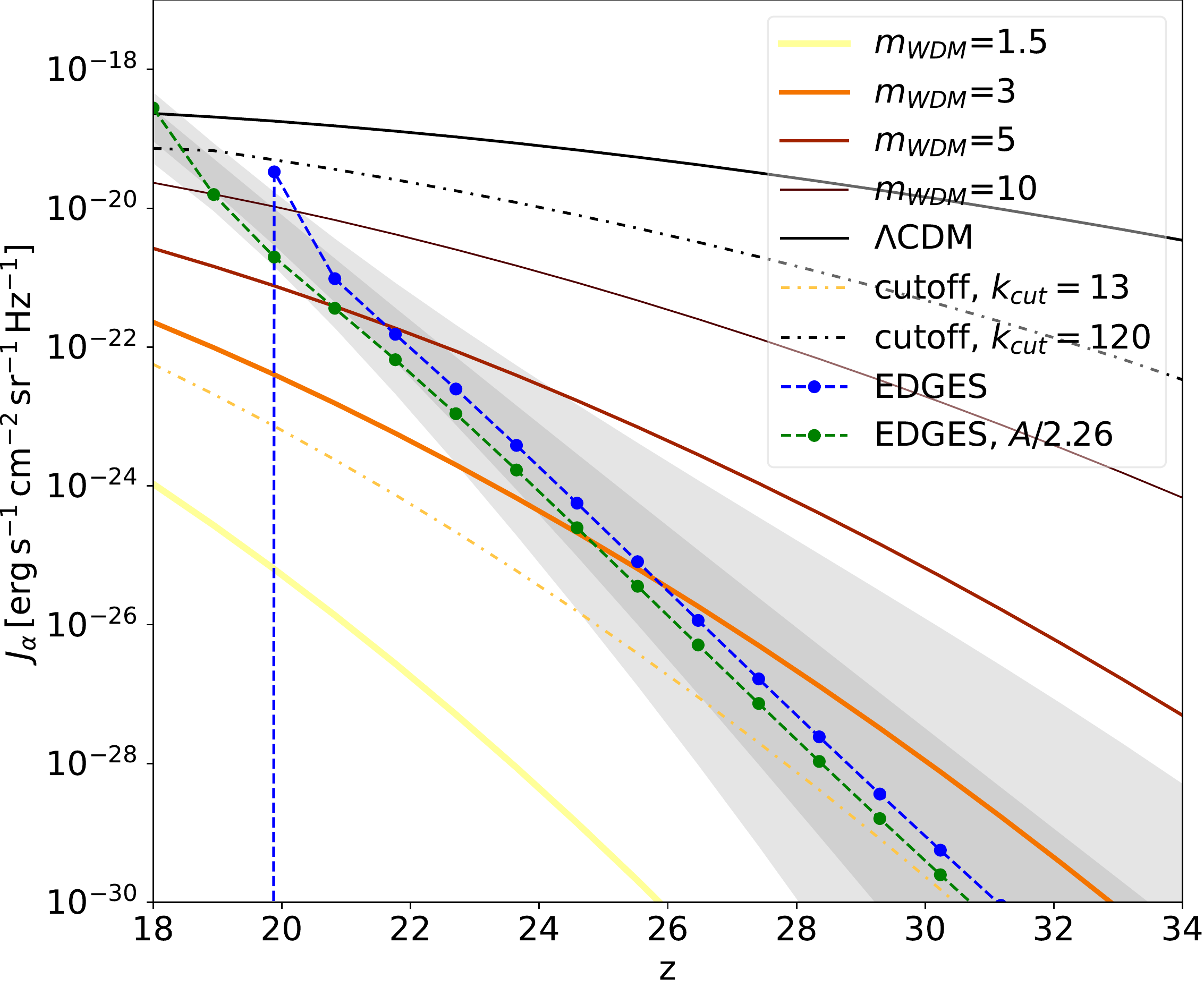}
    \caption{Resulting Ly$\alpha$ intensities for WDM models of different masses, $\Lambda$CDM (both solid lines), and $\Lambda$CDM with a cutoff (dash-dotted). For all these lines star formation efficiency is $f_\star=0.03$. The blue dashed line with data points shows the Ly$\alpha$ intensity calculated from EDGES best-fit parameters. If the evolution of $T_R$ and $T_K$ are unchanged, then at $z \sim 20$ spin temperature drops below $T_K$ causing a breakdown of the model (see also in figure~\ref{fig:t21comp}). Reducing the amplitude of the EDGES signal by roughly half to $A=0.23$ (green dashed line with datapoints), such that $T_S>T_K$ at all redshifts, then the resulting Ly$\alpha$ intensity changes only in amplitude, not slope. Grey-shaded areas give 1$\sigma$ and 2$\sigma$ uncertainty regions for the Ly$\alpha$ intensity derived from the EDGES signal.}
    \label{fig:sfrdcomp}
\end{figure}

\begin{figure}
    \centering
    \includegraphics[width=0.48\textwidth]{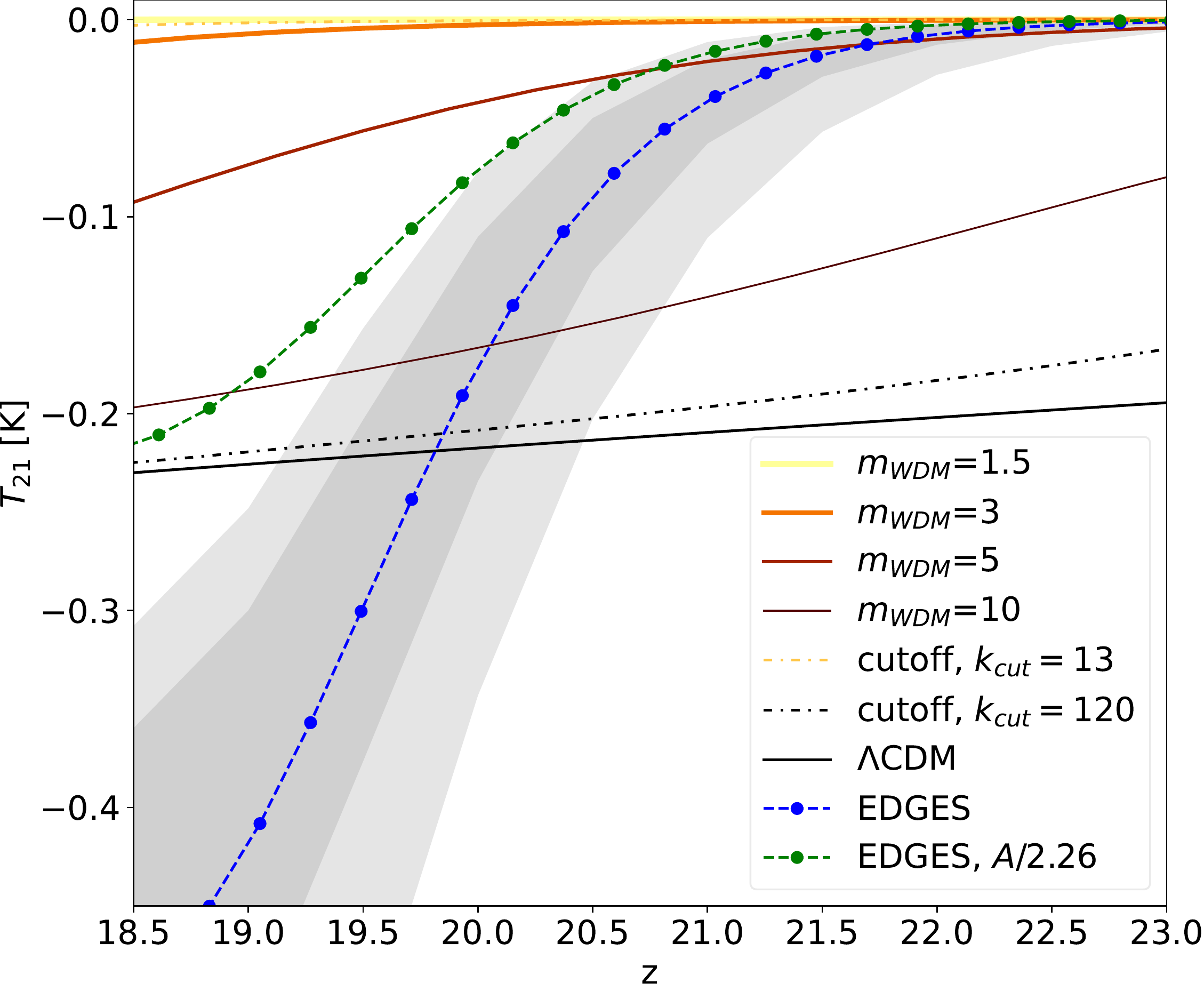}
    \caption{Comparison of antenna brightness temperature for WDM models with different masses, the EDGES best-fit line, and the same best-fit line but with reduced amplitude. Colours, line styles and model parameters match those in figure \ref{fig:sfrdcomp}. The $J_\alpha$ generated by $\Lambda$CDM is enough to couple the corresponding $T_{21}$ fully to $T_K$ very early on while low mass WDM models (and CDM models with low $k_{\mathrm{cut}}$) exhibit no absorption feature at those redshifts. Intermediary models have a more noticeable, though much too flat, absorption feature. The $\Lambda$CDM line also marks the lowest absorption amplitude possible with standard temperature histories and thus demonstrates the point at $z\sim 20$ where the EDGES signal becomes problematic. Grey-shaded areas give 1$\sigma$ and 2$\sigma$ uncertainty regions for the Ly$\alpha$ intensity derived from the EDGES signal.}
    \label{fig:t21comp}
\end{figure}

There is no consensus for the average emissivity of stars at high redshifts because the ratio of PopII to PopIII star formation is not certain. In this work we assume an ongoing star formation process that forms PopIII stars which emit radiation according to the emissivity model used by \cite{schauer_constraining_2019} in which the number of ionizing photons emitted per second by PopIII stars is $\dot{N}_{ion}=10^{48} \, \mathrm{s}^{-1} \mathrm{M}_\odot^{-1}$ and the lifetime of these stars is $t_\star = 3$ Myr. It is important to note that the choice of an emissivity model can have a large impact on the outcome. For example, the Ly$\alpha$ intensities calculated from emissivities detailed in \cite{pritchard_descending_2006} and \cite{schauer_constraining_2019} differ by roughly an order of magnitude. In principle any difference in the emissivity model can be compensated by changes in the values of $f_\star$ and $f_R$ during fitting and we find this to be true in our MCMC runs. The ongoing star formation model contains no feedback, so previous generations of PopIII stars have no effect on subsequent star formation. The SFRD is proportional to the fraction of mass in collapsed halos, $f_{\mathrm{coll}}(z)$, which is a simple integral over the HMF
\begin{equation}
    f_{\mathrm{coll}}(z) = \dfrac{1}{\rho_m^0} \int_{M_{\mathrm{min}}}^\infty \dfrac{\dif n}{\dif \ln M} \dif M,
\end{equation}
where $\rho^0_m$ is the mean matter density at $z=0$ and the lower bound for integration is set by the virial temperature and the type of cooling process, atomic \mbox{($T_{\mathrm{vir}}=10^4$ K)} or molecular ($T_{\mathrm{vir}}=10^3$ K) hydrogen, which is active \cite{barkana_beginning_2001} 
\begin{align}
    M_{\mathrm{min}} =& 10^8 \left( \dfrac{1+z}{10}\right)^{-3/2} \left( \dfrac{\mu}{0.6} \right)^{-3/2} \nonumber \\
    & \times  \left( \dfrac{T_{\mathrm{vir}}}{1.98\cdot 10^4 \mathrm{K}} \right)^{3/2} \left( \dfrac{\Omega_m}{\Omega_m^z} \dfrac{\Delta_c}{18\pi^2} \right)^{-1/2}. 
\end{align}
Here the last term reduces to 1 because at high redshifts $\Omega_m^z\approx\Omega_m$ and $\Delta_c\approx18\pi^2$~\cite{bryan_statistical_1998}. As we are looking at non-CDM models where small scale halos are suppressed, the precise choice of $T_{\mathrm{vir}}$, and therefore $M_{\mathrm{min}}$, is relatively unimportant because there are fewer halos near $M_{\mathrm{min}}$ compared to the case of CDM. 

Figure~\ref{fig:sfrdcomp} shows Ly$\alpha$ intensities for WDM, CDM, and CDM with a sharp cutoff. It also shows the $J_\alpha$ line calculated directly from the EDGES best-fit parameters (with no added cooling, heating, or radiation background) with 1$\sigma$ and 2$\sigma$ intervals and the same line with amplitude reduced such that $T_R>T_S>T_K$ at all redshifts.

Both non-CDM models converge onto the CDM model in the same manner when their respective cut-off scale nears the chosen virial temperature and also at lower redshifts. It is clear that no model is nearly sharp enough to approximate the EDGES best-fit result in the case of standard radiation and gas temperature histories. Thus, no pure DM model fits the onset of the EDGES signal.

The discrepancy is even more apparent in figure~\ref{fig:t21comp}, where we compare $T_{21}$ for the same models as in figure~\ref{fig:sfrdcomp}. No model provides an antenna brightness temperature deep or sharp enough to match the EDGES measurement. Figures~\ref{fig:sfrdcomp} and \ref{fig:t21comp} also illustrate the point at which the EDGES result runs into problems: at $z\sim20$ the $T_{21}$ of EDGES surpasses the value allowed by a standard radiation and gas temperature history, also pushing $J_\alpha$ to unphysical behaviour. To prevent this we add extra radio background into our model as explained in detail in the next section.


\subsection{Added radio background and X-ray heating} \label{ssec:radio_xray}


One can explain the large amplitude of the EDGES signal by adding an extra radio background at the transition frequency of the spin transition of neutral hydrogen. Also, the detection of an excess radio background by the ARCADE-2 collaboration seems to support the idea of extra background~\cite{fixsen_arcade_2011}. Following the reasoning of~\cite{chatterjee_ruling_2019}, we look at the case where extra radio background is tied to the star formation rate. Thus, the extra radiation at the HI spin transition frequency of 1420~MHz is given as \cite{ciardi_probing_2003}
\begin{equation}
    J_R = \left( \dfrac{1420}{150} \right)^{-0.7} \dfrac{c (1+z)^3}{4\pi} \int_z^\infty \epsilon_R (z^\prime) \dif t,
\end{equation}
where the first factor is from extrapolating the local radio-to-star-formation-rate relation~\cite{gurkan_lofarh-atlas_2018} \mbox{$L_R = f_R \cdot 10^{22} \dot{M}_\star$} to relevant frequencies with the spectral index $-0.7$. 
The radio emissivity is tied to the SFRD by
\begin{equation}
    \epsilon_R(z) = f_R \cdot 10^{22} \dfrac{\dot{\rho}_\star}{M_\odot \mathrm{yr}^{-1} \mathrm{Mpc}^{-3} } \mathrm{J} \, \mathrm{s}^{-1} \mathrm{Hz}^{-1} \mathrm{Mpc}^{-3}.
\end{equation}
In this model $f_R$ is a free parameter that, similarly to $f_\star$, determines the efficiency and cosmological spread of radiation production. 

The source of the extra radio background added in this way can be approximated as a black body, the temperature of which can be calculated from the Rayleigh-Jeans law. Then the new radiation temperature is the sum of the CMB and this extra radio source $T_{R, new}(z)=T_R(z) + T_{radio}(z)$. Mirocha \& Furlanetto 2019 found that in order to satisfy the ARCADE-2 excess at $z=0$ the production of extra radio background must be switched off by $z\sim 15$~\cite{mirocha_what_2019}. This limit is insignificant in our model since we concentrate on the high-$z$ end of the signal. Instead we illustrate the effect of cutting the radio production at a higher redshift, at $z\sim 20$. Once one cuts off the production the only further effect on the radio background is the adiabatic expansion of the Universe. Figure~\ref{fig:bestfits} shows the effects of such a treatment.

Though we focus on the onset of the EDGES signal we also apply a simple heating model to demonstrate how the whole signal shape for such a model might look like. For this we tie the heating of gas through X-ray luminosity to the SFRD, analogous to the radio background and Ly$\alpha$ intensity. Following~\cite{furlanetto_effects_2006} and \cite{chatterjee_ruling_2019} the evolution of $T_K$ can be expressed as
\begin{equation}
    \dfrac{\dif T_K}{\dif z} = \dfrac{2T_K}{1+z} - \dfrac{2}{3H(z)(1+z)} \dfrac{\epsilon_X }{k_B n_{\mathrm{gas}}(z)},
\end{equation}
where the first term is adiabatic cooling due to expansion of the Universe, the second is heating from X-ray photons, $\epsilon_X$ is the energy added by X-ray heating, $k_B$ is Boltzmann's constant, and $n_{\mathrm{gas}}$ the average number density of gas particles at any given redshift. The added heating can be expressed as \cite{chatterjee_ruling_2019}
\begin{equation}
    \epsilon_X = 3.4\cdot 10^{33} f_X \, \dot{\rho}_\star \, \, \mathrm{ J \, s^{-1} \, Mpc^{-3}},
\end{equation}
with $f_X$ as a free parameter like $f_R$ and $f_\star$. For simplicity we leave aside other heating sources like Compton or Ly$\alpha$ heating. Figure~\ref{fig:bestfits} shows the signal shapes this simple heating model provides.


\section{Results}\label{sec:results}

\begin{figure*}
    \centering
    \includegraphics[width=0.75\textwidth]{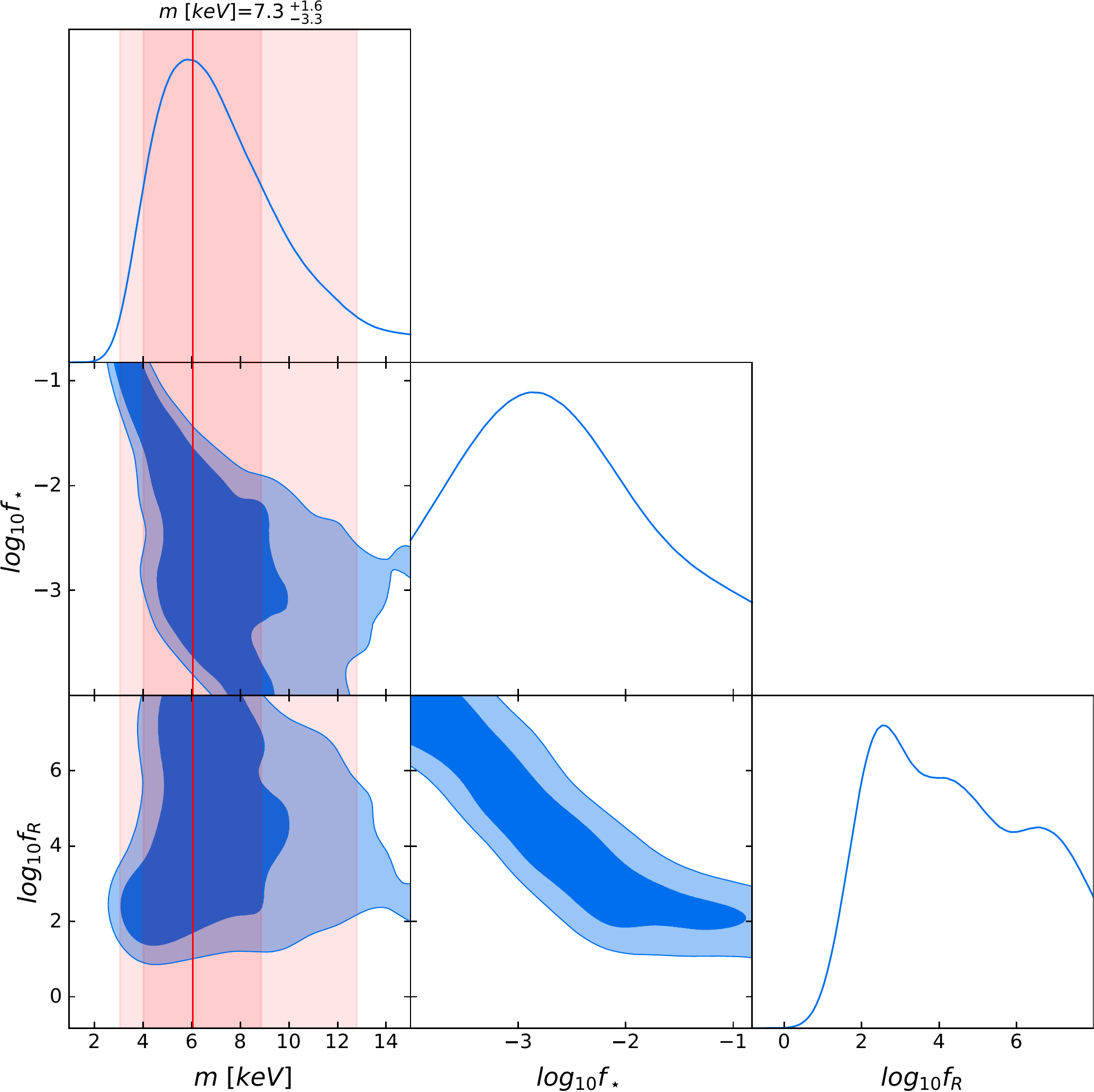}
    \caption{Posterior parameter distributions from MCMC calculations. The mean value for $m$ with 1$\sigma$ error bounds is also shown. The red marker shows the location of the best fitting mass value and shaded areas show $1\sigma$ and $2\sigma$ regions around the mean. Calculations showed that $f_\star$ and $f_R$ are somewhat degenerate; while the posterior distributions for these parameters exhibit clear maxima (driven by our choice of priors), there are no corresponding maxima in their likelihood values. Therefore there are no strongly preferred values for them. This plot is made with the package getdist~\cite{lewis_getdist_2019}.}
    \label{fig:mcmcdists}
\end{figure*}

The parameter distributions resulting from our MCMC run for thermal WDM are shown in figure \ref{fig:mcmcdists} along with the statistical average, best-fit value, and 1$\sigma$ and 2$\sigma$ confidence intervals for WDM mass. As expected, the parameters are not wholly independent; there is a strong correlation between $f_\star$ and $f_R$ (the values of which are also model dependent) and a somewhat weaker correlation between $f_\star$ and particle mass. As such we are mainly interested in the posterior distribution for particle mass with the ranges for $f_\star$ and $f_R$ chosen such that they do not set hard limits to the mass range. An exception is the physically motivated upper bound of $f_\star<0.15$ mentioned in section \ref{ssec:sfrd}. 

The resulting posterior distributions exhibit a clear maximum in mass which must be attributed to the rate of structure formation in the corresponding model. Similar maxima, while present in the posterior distributions of $f_\star$ and $f_R$, are not reflected in their likelihood ($\chi^2)$ values and are mostly driven by the assumed prior ranges for those parameters. Thus, the most important parameter in the model is particle mass -- the others being degenerate enough that their specific values are relatively unimportant. 

\begin{figure}
    \centering
    \includegraphics[width=0.48\textwidth]{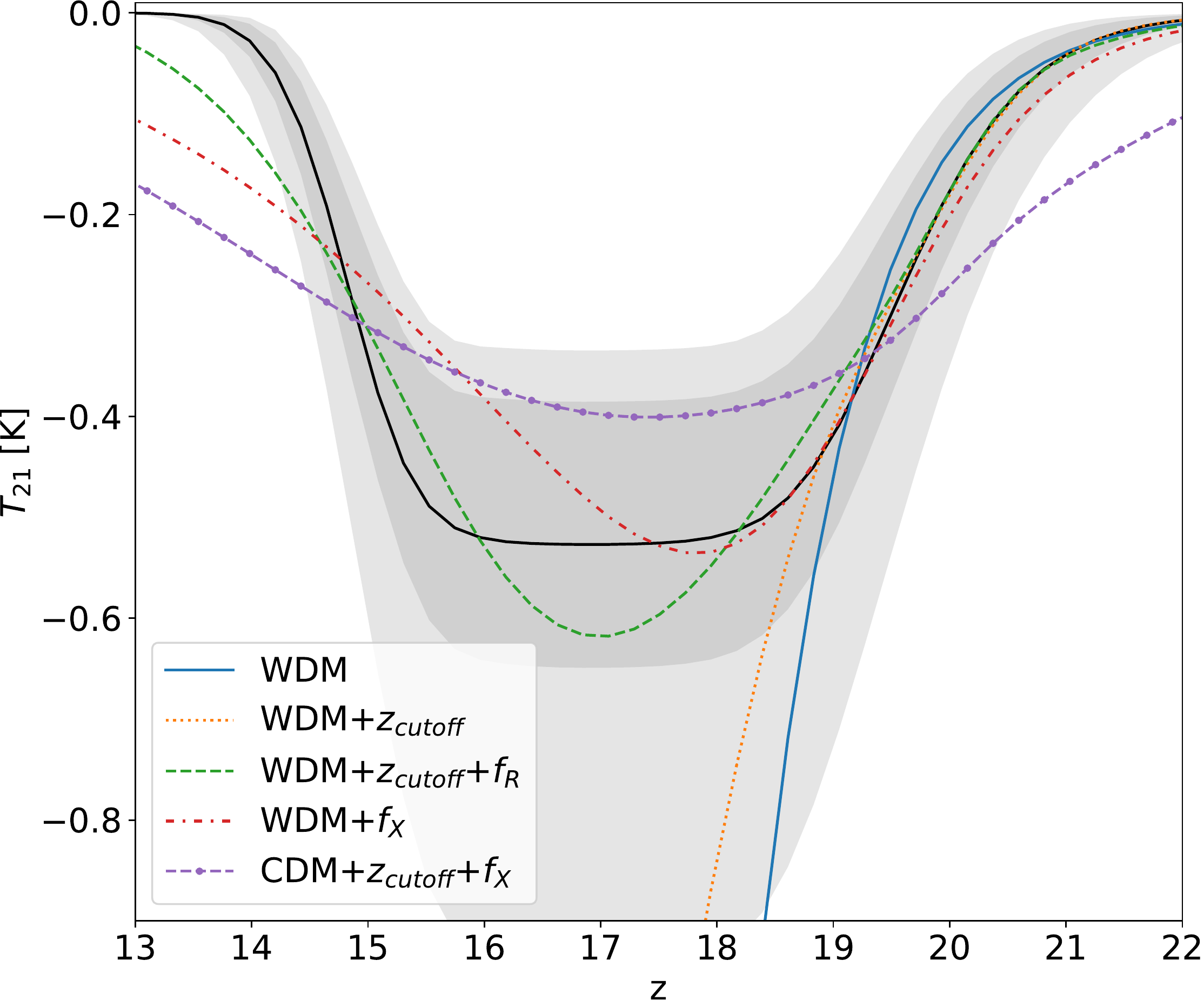}
    \caption{Comparison of the best fitting models with the parameter values listed in table \ref{tab:pars_bestfit} along with the EDGES result and its 1$\sigma$ and 2$\sigma$ uncertainty regions. The first two lines (solid blue and dotted orange) are fitted only to the onset of the EDGES signal. Others are fitted to the full signal.}  
    \label{fig:bestfits}
\end{figure}

As figure \ref{fig:bestfits} demonstrates, the base WDM model (solid blue line), where collapsing structures produce Ly$\alpha$ radiation and extra radio background, can fit the EDGES onset relatively well. To illustrate the effects that additional parameters have on a simple WDM model, we show lines with added radio cutoff $z_{\mathrm{cutoff}}$ and/or X-ray heating $f_X$. When radio production is cut at $z\sim 20$ (dotted orange) the fit improves and allows for a lower particle mass. If there is no heating included, the growth of radiation temperature is not counteracted by a growth of gas temperature and the fall of $T_{21}$ is not bounded. With added heating (green dashed and red dash-dotted), WDM can fit the onset but not the flat bottom or sharp end of the signal. A model with both heating and a radio cutoff (green dashed) produces the best fit for the overall signal shape.

\begin{table} 
\centering
{ \setlength{\tabcolsep}{0.6em} \setlength{\extrarowheight}{0.2em}
\begin{tabular} { l | c c c c | c} 
 Parameter & \multicolumn{4}{c}{ Thermal WDM} & CDM \\ 
\hline 
{\boldmath$m_x \, [\mathrm{keV}] $} & 6.05  & 4.56   & 4.26  & 3.7   & - \\
{\boldmath$\log_{10}f_\star  $}     & -1.93 & -2.19  & -1.32 & -1.11 & -3.41 \\
{\boldmath$\log_{10}f_R   $}        & 2.65  & 4.81   & 2.71  & 2.43  & 3.26 \\
{\boldmath$z_{\mathrm{cutoff}}  $}           & -     & 20.15  & -     & 19.7  & 19.29 \\
{\boldmath$\log_{10}f_X $}          & -     & -      & 4.39  & 2.38  & 2.9 \\
\end{tabular} 
\caption{ Best fitting parameter values for CDM and different WDM models.} 
\label{tab:pars_bestfit}}
\end{table}

No CDM model is able to reproduce the rapid onset of the EDGES signal. The CDM model plotted in figure \ref{fig:bestfits} (purple dashed line with dots) is representative of the best-fitting CDM curve with any set of parameters; without $z_{\mathrm{cutoff}}$ the best fit is more symmetrical, without $f_X$ it simply does not deviate from the original shallow downward slope.

Thus, the EDGES signal (or more accurately its rapid onset) sets both upper and lower limits for small scale suppression. The posterior average WDM particle mass (figure \ref{fig:mcmcdists}) is $7.3^{+1.6}_{-3.3}$ keV, meaning that at masses $\lesssim 4$ keV the suppression of small scale structure is too strong (structures form too late) and at masses $\gtrsim 9$ keV the suppression is not strong enough (structures form too soon). In table \ref{tab:mass_translate} we translate this mass range into their equivalent FDM and Dodelson-Widrow sterile neutrino model values. Comparing this mass bound to Ly$\alpha$ production in figure \ref{fig:sfrdcomp} (which is tied to SFRD) we see that EDGES requires the rate of structure growth to advance within a relatively narrow range.
\begin{table}  
\centering
{ \setlength{\tabcolsep}{0.4em} \setlength{\extrarowheight}{0.4em}
\begin{tabular} { l | c c c }
 & WDM & FDM & Sterile neutrino \\ 
\hline 
Mass & $7.3^{+1.6}_{-3.3}$ keV & $2.2^{+1.4}_{-1.7} \cdot 10^{-20}$ eV & $63^{+19}_{-35}$ keV  \\ \end{tabular} }
\caption{ Mass values that are compatible with the rapid onset of EDGES. Masses for WDM are the posterior mean from our MCMC calculations along with 1$\sigma$ error bounds. The FDM mass range is calculated from WDM by equation \eqref{eq:wdm_to_fdm} and the sterile neutrino range by equation \eqref{eq:wdm_to_nu}.}
\label{tab:mass_translate}
\end{table}

The mass range we find for WDM is compatible with the current lower limit of $m_x > 5.3$ keV~\cite{irsic_new_2018}. For FDM the bounds from density profiles of dwarf galaxies cover a wide range and are not necessarily in agreement with each other or this work (see \emph{e.g.}~\cite{chen_jeans_2017, gonzalez-morales_unbiased_2017, zoutendijk_muse-faint_2021}). The Ly$\alpha$ forest provides a lower bound of $m_{\mathrm{FDM}} > 2\cdot 10^{-21}$ eV~\cite{irsic_first_2017} which is compatible with this work. Sterile neutrino models have an additional degree of freedom in the form of the mixing angle $\theta$. Current limits for $m_\nu$ and $\theta$ can be found in~\cite{abazajian_neutrinos_2021}. The lower limit for WDM translates through equation \eqref{eq:wdm_to_nu} to $m_\nu > 41$ keV. The recent sterile neutrino candidate of mass 7.1 keV~\cite{bulbul_detection_2014} lies far from the range found in this work. 

We find that the \emph{ad hoc} model of CDM with a cutoff in its power spectrum results in SFRDs qualitatively identical to those from the physically motivated WDM model (see figure \ref{fig:sfrdcomp}). The resulting mass range corresponds to cutoffs at wavenumbers $44^{+11}_{-22} \, h/\mathrm{Mpc}$. A phenomenologically similar model was used in~\cite{kaurov_implication_2018} for fitting the overall shape of the EDGES signal, although with a reduced amplitude, to argue that most of the star formation must have occurred in very massive halos with $M \gtrsim 10^9 \, \mathrm{M}_\odot$. At redshift $z\sim 20$ this corresponds to wavenumbers $k\lesssim 15 \, h/\mathrm{Mpc}$ (WDM with $m_x \lesssim 2.5$ keV) and therefore such models are disfavored by our results. Finally, we note that a cut in the production of radio background at $z\sim 20$ may improve model fit quality by allowing for a flatter bottom for the $T_{21}$ signal of a given model.


\section{Conclusions and discussion}\label{sec:conclusions}



In this work we calculated the 21-cm antenna brightness temperature for a base WDM model and used the onset of the EDGES signal to set bounds on the allowed mass of the WDM particle. Our model included a sharp-$k$ window function for the HMF, which has been found to be accurate at high redshifts~\cite{schneider_halo_2013}. The production of Ly$\alpha$ radiation and extra radio background was tied to the SFRD of the model.

The EDGES signal is a unique messenger, currently the only signal originating from the cosmic dawn. Probes at lower redshifts mostly allow one to put upper limits on suppression of small-scale power (lower limits on non-CDM particle mass) because at low redshifts higher mass non-CDM models (for example $m_x \gtrsim 10$) are nearly indistinguishable from CDM at typically accessible scales. At the redshifts of cosmic dawn ($z\sim 20$) the divergence between models is greater. This makes the rapid onset of EDGES, which can be used to set bounds on the speed of structure formation and thus set both upper and lower limits for suppression of small scale structure, very significant.

We find that the onset of the EDGES signal limits the mass of thermal WDM particles to $7.3^{+1.6}_{-3.3}$ keV within 1$\sigma$, mass of FDM particles to $2.2^{+1.4}_{-1.7} \cdot 10^{-20}$ eV, and mass of Dodelson-Widrow sterile neutrinos to $63^{+19}_{-35}$ keV. These ranges are compatible with current limits set for WDM and FDM by Ly$\alpha$ forest observations.

It is important to note that the EDGES signal has two unexpected characteristics which require explanation: its high amplitude and its rapid onset. The high amplitude can be explained by either cooling of baryon gas (lowering $T_K$) or generating extra radio background (raising $T_R$). Even a simple scaling of either of these quantities is enough to achieve this. As for the onset, one must either modify the \emph{slope} of $T_K$ or $T_R$ or otherwise modify star formation to occur much more rapidly. It is difficult to achieve a cooling effect that is sharp and occurs exactly at relevant redshifts (an exception is for example the cooling from primordial magnetic fields \cite{natwariya_edges_2020}) and this is evidenced by how works which add cooling (by way of baryon-DM interaction~\cite{barkana_possible_2018,liu_too_2018,munoz_insights_2018}) ignore the sharpness of the EDGES signal. As demonstrated in this work and in the references~\cite{ewall-wice_modeling_2018, chatterjee_ruling_2019}, an extra radio background generated in tandem with the onset of Ly$\alpha$ sources can more easily achieve the sharpness of the observed EDGES signal.

We look forward to future measurements of the global 21-cm signal or upcoming measurements with interferometers that will help to solidify the result and/or set more accurate bounds for feasible non-CDM models.

\begin{acknowledgments}
We thank Raul Monsalve for EDGES related clarifications and our referee for comments and suggestion which helped to improve the paper. This work was supported by the European Regional Development Fund through the CoE program grant TK133, the Mobilitas Pluss grants MOBTT5, and by the Estonian Research Council grants PRG434 and PRG803.
\end{acknowledgments}

\bibliographystyle{apsrev4-2}
\bibliography{refs.bib}

\end{document}